\documentclass[twocolumn,floatfix,pra,aps,showpacs,superscriptaddress,floatfix,footinbib]{revtex4}
\usepackage{epsfig,graphicx,tabularx}
\usepackage{amsmath}
\usepackage{amssymb}
\usepackage{color,braket}
\usepackage{url}

\definecolor{myblue}{RGB}{0,128,255}
\definecolor{darkblue}{RGB}{0,00,255}
\definecolor{mypurple}{RGB}{255,0,255}
\definecolor{lightred}{RGB}{255,0,128}

\newcommand{\mb}{\mathbf}
\newcommand{\be}{\begin{equation}}
\newcommand{\ee}{\end{equation}}
\newcommand{\sm}[1]{\left|#1\right|^2}
\newcommand{\dd}{\mathrm{d}}
\newcommand{\ii}{\mathrm{i}}

\newcommand{\rtfsq}{{R_{\mathrm{TF}}}^2}
\newcommand{\rtf}{R_{\mathrm{TF}}}
\newcommand{\oho}{\omega_{ho}}
\newcommand{\ohosq}{{\omega_{ho}}^2}

\begin{document}

\title{Spin dipole oscillation and relaxation of coherently coupled Bose-Einstein condensates}

\author{A. Sartori}
\affiliation{INO-CNR BEC Center and Dipartimento di Fisica, Universit\`a di Trento, 38123 Povo, Italy}
\author{J. Marino}
\affiliation{Institute of Theoretical Physics, TU Dresden, D-01062 Dresden, Germany}
\author{S. Stringari}
\affiliation{INO-CNR BEC Center and Dipartimento di Fisica, Universit\`a di Trento, 38123 Povo, Italy}
\author{A. Recati}\email{recati@science.unitn.it}
\affiliation{INO-CNR BEC Center and Dipartimento di Fisica, Universit\`a di Trento, 38123 Povo, Italy}
\affiliation{Technische Universit\"at M\"unchen,  James-Franck-Strasse 1, 85748 Garching, Germany}

\begin{abstract}
We study the static and the dynamic response of coherently coupled two component  Bose-Einstein condensates due to a spin-dipole perturbation.  
The static dipole susceptibility is determined and it is shown to be  a key quantity to identify the second order ferromagnetic transition occurring at large inter-species interaction. The dynamics, which is obtained by quenching the spin-dipole perturbation, is very much affected by the system being paramagnetic or ferromagnetic and by the correlation between the motional and the internal degrees of freedom.
In the paramagnetic phase the gas exhibits well defined out-of-phase dipole oscillations, whose frequency can be related to the susceptibility of the system using a sum rule approach. In particular in the interaction $SU(2)$ symmetric case, i.e., all the two-body interactions are the same, the external dipole oscillation coincides with the internal Rabi flipping frequency.
In the ferromagnetic case, where linear response theory in not applicable, the system show highly non linear dynamics. In particular we observe phenomena related to ground state selection: the gas, initially trapped in a domain wall configuration, reaches a final state corresponding to the magnetic ground state plus small density ripples.
Interestingly the time during which the gas is unable to escape from its initial configuration is found to be  proportional to the square root of the wall surface tension.
\end{abstract}

\maketitle

\section{Introduction} 

Ultra-cold gases allow the realisations of multi-component Bose-Einstein condensates (BECs). The latter are novel systems, whose behaviour is very different with respect to the one of single component BEC.
In particular they show different zero-temperature phases, each described by a proper vector order parameter. 
The possibility of tuning a number of system's parameters, in particular the interaction strength through Feshbach resonances, make such systems ideal to study the structure of the various phases and the nature of the phase transitions.

One of the easiest, but still intriguing realisation is represented by a two-component BEC, also known as a spinor condensate.  
Spinor condensates allow to address many interesting phenomena from Andreev-Bashkin effect \cite{AndreevBashkin,AndreevBEC} and fast decay of persistent currents \cite{ZoranSC}, to (internal) Josephson effect \cite{ChapmanJJ,Zibold2010}, or Schr\"odinger-cat- and twin-Fock-like states \cite{Cirac1998,OberthalerTwin}, from dimerised vortices \cite{SonDVortex,UedaDVortex,SpiralsPRA2004}, to the study of quenching in classical bifurcations \cite{Lee2004,SabbatiniZurekPRL,DallaTorreInst}. They represent also the basis for most of the recent realisation of artificial gauges in cold gases\cite{DalibardGaugeRMP} .

In this paper we specifically consider a zero-temperature trapped two-component BEC with an external field that drives the population transfer (spin-flipping) between the two atomic levels (see Sec. \ref{sec:GP}) forming the condensate. It is common to refer to the interconversion term as a Rabi coupling. The system is indeed a generalisation to non-linear  atom optics of the well-known linear Rabi problem and in general is an extension of quantum optics concepts to condensates \cite{Blakie1999,Nicklas2011}. 
It is the interplay between the intra- and inter-species two-body interaction strengths and the Rabi coupling strength that makes the physics of the system very reach. The Rabi coupling -- which acts as a $\sigma_x$ operators on each atom -- tries to create an equal superposition of the two possible internal levels.
On the other hand differences in the three possible atom-atom interaction strengths tries to favour a situation where the population of the two internal levels is unbalanced. It turns out that the system exhibits a second order phase transition, a classical bifurcation at the mean-field level (see, e.g., \cite{Goldstein1997,Tommasini2003} and in particular the experiment \cite{Zibold2010}), which is analogous to the mean-field ferromagnetic transition of the Ising model in transverse field.
Moreover if the two component feel a different external potentials the internal and external degrees of freedom are inseparable leading to interesting spin-orbit coupled dynamics as it has already been shown some years ago in \cite{RabiCornell99,RabiCornell2000}. 

In the following we show that the static and the dynamic response to an out-of-phase (spin) dipole perturbation is very rich and captures many relevant phenomena related to the paramagnetic-ferromagnetic-like phase transition of the system.  
A relative component perturbation is accessible in cold gases by applying different trapping potentials for different atomic internal levels. The spin dipole configuration is realised by applying trapping potentials that have the same shape, but that are displaced for the two components of the gas. The dynamics is obtained by monitoring the gas after the displacement is suddenly set to zero. 
Notice that in \cite{RabiCornell99,RabiCornell2000} a similar situation has already been realised, where instead the external potentials were held fixed and the Rabi coupling suddenly turned on.  

The main results of our analysis can be summarised as follow: 

\noindent (i) in the region before the bifurcation occurs, i.e., in the paramagnetic phase the system exhibits well defined out-of-phase oscillations around the equilibrium position in the new trapping potential. The oscillation frequency is in good agreement with a sum-rule approach calculation. The latter allows us to identify the main quantity determining the spin dipole mode frequency and its relation with the susceptibility of the system. In the case of equal interaction strength the sum rule give an exact result. The latter is practically twice the Rabi coupling, i.e., the main contribution is not proportional, as it is usually the case, to the harmonic trapping frequency. This effect can be traced back in the modification of the $f$-sum rule, which is eventually due to the absence of relative number conservation.

\noindent (ii) in the broken $\mathbb{Z}_2$, i.e., ferromagnetic phase the situation is very different. The response of the system to the spin dipole perturbation is not linear and therefore the initial state in the displaced potentials is far from the equilibrium state when the potentials are the same. In particular the initial configuration shows a polarisation domain wall at the center of the cloud, but zero global polarisation, while in the new equilibrium state it will show a symmetric structure with a global polarisation. The dynamics is highly non-linear. After a certain period -- in which the system is trapped in the domain wall configuration -- the cloud is able to reach quickly the new equilibrium by spontaneously selecting one of the two possible polarisation.  The excess energy of the initial configuration gives rise to small ripples in the cloud.
 
It is worth mentioning here that the very same mean-field description we use in the following (see Sec. \ref{sec:GP}) can be applied, in certain regimes, to describe polariton systems where the role of the polarisation is relevant (see e.g., \cite{polariton1,polariton2}), as well as some properties of type-1.5 superconductors (see \cite{GLtype1.5} and reference therein).

The paper is organised as it follows. In Sec. \ref{sec:GP} we introduce the system and its description in terms of two coupled Gross-Pitaevskii (GP) equations. We revisit the emergence of a paramagnetic/ferromagnetic like transition and the effect of the external harmonic trapping potential. In Sec. \ref{sec:susc} we study the effect of a spin-dependent potential and the role of the spin-dipole susceptibility.  The latter is shown to bear a clear signature of the phase transition. Then, we address the problem of the dynamics of the spin-dipole mode both in the para- and in the ferromagnetic phase. In the former case (Sec. \ref{sec:para}) linear response theory combined with a sum-rule approach provides  an accurate estimate of the spin-dipole mode frequency, which well compares  with the numerical solution of the GP equation. In the ferromagnetic case (Sec. \ref{sec:ferro}) we show that the system exhibits ground state selection, after a waiting time in which the system is unable to leave the initial domain wall configuration. We found phenomenologically that this characteristic time is proportional to the square root of the domain wall energy. 


\section{Gross-Pitaevskii equation for coherently coupled BECs}\label{sec:GP}

We consider an atomic Bose gas at zero temperature, where each atom of mass $m$ has two internal levels $|a\rangle$ and $|b\rangle$. The latter are typically magnetically trappable hyperfine levels of $^{87}$Rb, like $|a\rangle=|F=1, m_F=-1\rangle$ ($|F=1, m_F=1\rangle$) and $|b\rangle=|F=2, m_F=1\rangle$ ($|F=2, m_F=-1\rangle$). An external field is applied that coupled the $|a\rangle$ to the $|b\rangle$ state via usually a two-photon transition, characterised by a Rabi splitting $\Omega$.  
At the densities of ultra-cold gases the atomic interactions are simply described by a contact potential with a strength proportional to the $s$-wave scattering length. For a spinor system three scattering lengths, $a_{aa}$, $a_{bb}$ and $a_{ab}$ are present describing the intra- and the inter-species collisions, respectively. Finally the condensed phase for a two-component Bose gas is described by a complex spinor order parameter $(\psi_a({\mb r},t),\psi_b({\mb r},t))$, where $\psi_i$, $i\in\{a,b\}$ is the wave function macroscopically occupied by atoms in the internal state $|i\rangle$. The latter is normalised to the total number of atoms $N_i$ in the state $|i\rangle$. 
The dynamics of the order parameter is determined by coupled Gross-Pitaevskii equations \cite{SandroLevBook,Blakie1999}
\begin{align} \label{eq:tGP}
\ii\hbar\frac{\partial}{\partial t}\psi_a &=\! \left[ -\frac{\hbar^2\nabla^2}{2m} + V_a + g_a\!\sm{\psi_a} + g_{ab}\!\sm{\psi_b} \right]\!\psi_a + \Omega\psi_b, \\
\ii\hbar\frac{\partial}{\partial t}\psi_b &=\! \left[ -\frac{\hbar^2\nabla^2}{2m} + V_b + g_b\!\sm{\psi_b} + g_{ab}\!\sm{\psi_a} \right]\!\psi_b + \Omega^*\psi_a,\nonumber
\end{align}
where the couplings $g_i$, with $i\in \{a,b,ab\}$, are the intra- and interspecies atomic interaction strengths and are given by $g_i\propto a_i$ \cite{SandroLevBook}, and $V_a$ and $V_b$ are the external trapping potentials. We consider the confinement to be harmonic, which is the most relevant and typical experimental situation.  In the following, if not differently specified, we consider $g_a=g_b\equiv g$.
Due to the presence of the Rabi coupling only the total number of atoms $N=N_a+N_b$ is conserved, but not its polarisation $P=(N_a-N_b)/N$. The (gauge) symmetry of the system is therefore reduced from $U(1)\times U(1)$ to $U(1)\times \mathbb{Z}_2$, leading for an homogeneous condensate to a gapless density or in-phase mode -- Goldstone mode of the broken $U(1)$ symmetry -- and a gapped spin or out-of phase mode (see, e.g., \cite{Goldstein1997,Tommasini2003}). Depending on the interaction strengths and the Rabi coupling the ground state can also spontaneously breaks the $\mathbb{Z}_2$ symmetry leading to $P\neq 0$.

\begin{figure}[t]
\centering
\includegraphics{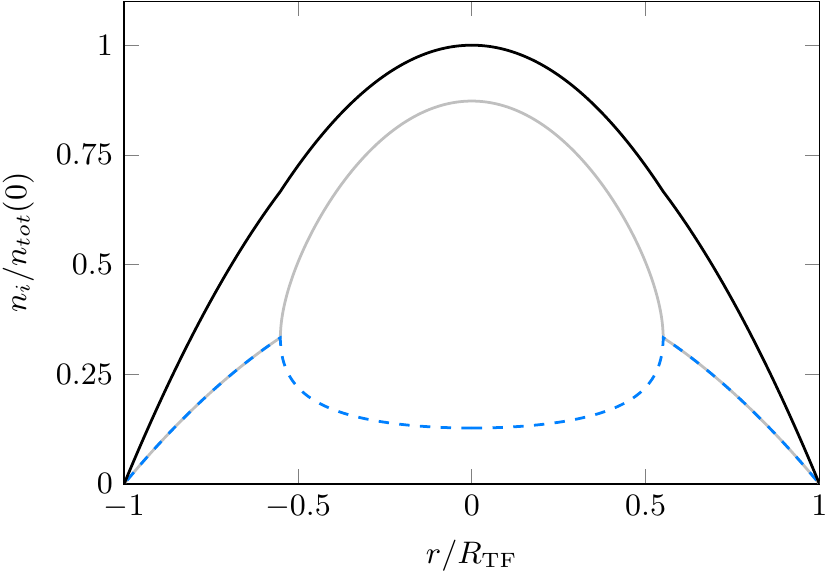}
\caption{Density profiles within Thomas-Fermi approximation for an harmonic trapping: $n_a$ (dashed blue), $n_b$ (light grey) and $n_a+n_b$ (black) for $g_{ab}/g=1.3$ and $\Omega/\mu=0.1$.}\label{fig:phase}
\end{figure}

In order to describe the ground state we write as usual the condensate wave function as density and phase $\psi_i=\sqrt{n_i}e^{i\phi_i}$ and use local density approximation (LDA), i.e., neglecting the gradient term, also known as quantum pressure, in Eqs. (\ref{eq:tGP}). The time derivative on the LHS of  Eqs. (\ref{eq:tGP}) is replaced by the chemical potential $\mu$, whose value will be fixed by requiring a total number of particle $N$. Notice that in absence of $\Omega$ one can have two different chemical potentials reflecting that also $P$ is fixed. The Rabi coupling originates a term of the form $\Omega\cos(\phi_a-\phi_b)$ for the energy. Without any loss of generality we also assumed $\Omega$ to be real and positive, which fixes the phases in the ground state to satisfy $\phi_-=\phi_a-\phi_b=\pi$. Finally one finds that the densities of the two component obey the relations   (see, e.g. the review \cite{marta2013} and reference therein) 

\begin{align}
\left( g - g_{ab} + \frac{\Omega}{\sqrt{n_an_b}} \right) \left( n_a - n_b \right) &= V_b-V_a,\label{eq:gs1}\\
\left( g + g_{ab} - \frac{\Omega}{\sqrt{n_an_b}} \right) \left( n_a + n_b \right) &= 2\mu - \left( V_b+V_a \right),\label{eq:gs2}
\end{align}

For the sake of simplicity and clarity, we consider a mean-field one-dimensional situation. The latter is experimentally realised by making two of the trapping frequencies strong enough in order for the motion along such directions to be frozen. The coupling constants are in this case renormalised and can be simply related to the scattering length and the trapping transverse frequency $\omega_\perp$ by $g_i=2\hbar\omega_\perp a_i$ for $i\in\{a,b,ab\}$.    
It is worth noticing that our results do not qualitatively change in the two- or three-dimensional case.  

From Eq. (\ref{eq:gs2}) is it clear that, for equal potentials, $V_a=V_b$, the system can sustain a finite polarisation only if $g_{ab}$ is sufficiently large. 
In that case it turns out that the $P\neq 0$ states are the ground states of the system. Notice that both the critical value of $g_{ab}$ and $P$ are density dependent. 
It is easy to find that the points $X_P$ at which the polarised phase can exist is fixed by the condition \be g_{ab}>g+2\Omega/n(X_P)\label{XP}\ee with $n(x)=n_a(x)+n_b(x)$ the total local density. Since in the harmonic trap the density decreases going outward from the trap center the system can exhibit two different regions: unpolarised tails with $n_a=n_b$ and a polarised core with $n_a\ne n_b$. Clearly, if the condition Eq. (\ref{XP}) is not satisfied at the center of the trap, where the total density is maximum, then the whole system is unpolarised.  This allows us to introduce a critical value of Rabi coupling defined by
\begin{equation}
\Omega_{cr}=\frac{1}{2}n(x=0)(g_{ab}-g)
\label{omegacr}
\end{equation}
For values $\Omega \ge \Omega_{cr}$ the system is unpolarised everywhere. Writing $V_a=V_b=m\omega_{ho}^2 x^2/2$, the density profile $n_a=n_b$ is easily obtained from Eq.(\ref{eq:gs2}):
\be
n_{a,b}(x)\equiv n_0(x) = \frac{\mu+\Omega}{g+g_{ab}}\left(1-\frac{x^2}{\rtfsq}\right),
\ee
where we introduced the so-called Thomas-Fermi radius  $\rtfsq=2(\mu+\Omega)/(m{\omega_{ho}}^2)$ \cite{SandroLevBook} and the chemical potential can be written as
\be
\mu = \left[\frac{3}{8}N(g+g_{ab})\right]^{2/3}\left(\frac{m\omega_{ho}}{2}\right)^{1/3}-\Omega.
\ee
 
In the case  $\Omega < \Omega_{cr}$ a typical configuration within LDA is shown in Fig. \ref{fig:phase}.

Let us here briefly remind that for a Bose-Bose mixture in the absence of Rabi coupling ($\Omega=0$), where the  relative particle number can be chosen at will, the situation is  very different. In that case there exists a first order phase transition to a phase separated state once $g_{ab}>g$ and the system in the trap is formed by two distinct region of only one of the two component of the gas (see for a detailed discussion , e.g. \cite{HoShenoyMix,SvidChuiPS,ModugnoRiboli}). An example of the structure for an equal number of atoms in both hyperfine levels is shown in Fig. \ref{fig:dipgs} (a1). 


\section{Static dipole polarizability}\label{sec:susc}

In this Section we  calculate the static response of a trapped spinor gas to a spin-dipole perturbation.  
A spin-dipole perturbation corresponds to a shift of the harmonic traps for the two components by a quantity $d\ll x_{ho}$ with $x_{ho} = \sqrt{\hbar/(m\omega_{ho})}$,
\begin{align}
\nonumber V_{a,b}&=\frac{1}{2}m{\omega_{ho}}^2\left(x\pm d\right)^2 \\
&= \frac{1}{2} m\omega_{ho}^2x^2 \pm m\omega_{ho}^2 x d+\mathrm{O}(d^2)\label{eq:diptrap},
\end{align}
where the plus sign is for particles of component $a$ and the minus one for those of component $b$. In the case of hyperfine atomic levels the displacement  can be realised by adding a magnetic field gradient to the harmonic potential. 

\begin{figure*}[t]
\centering
\includegraphics{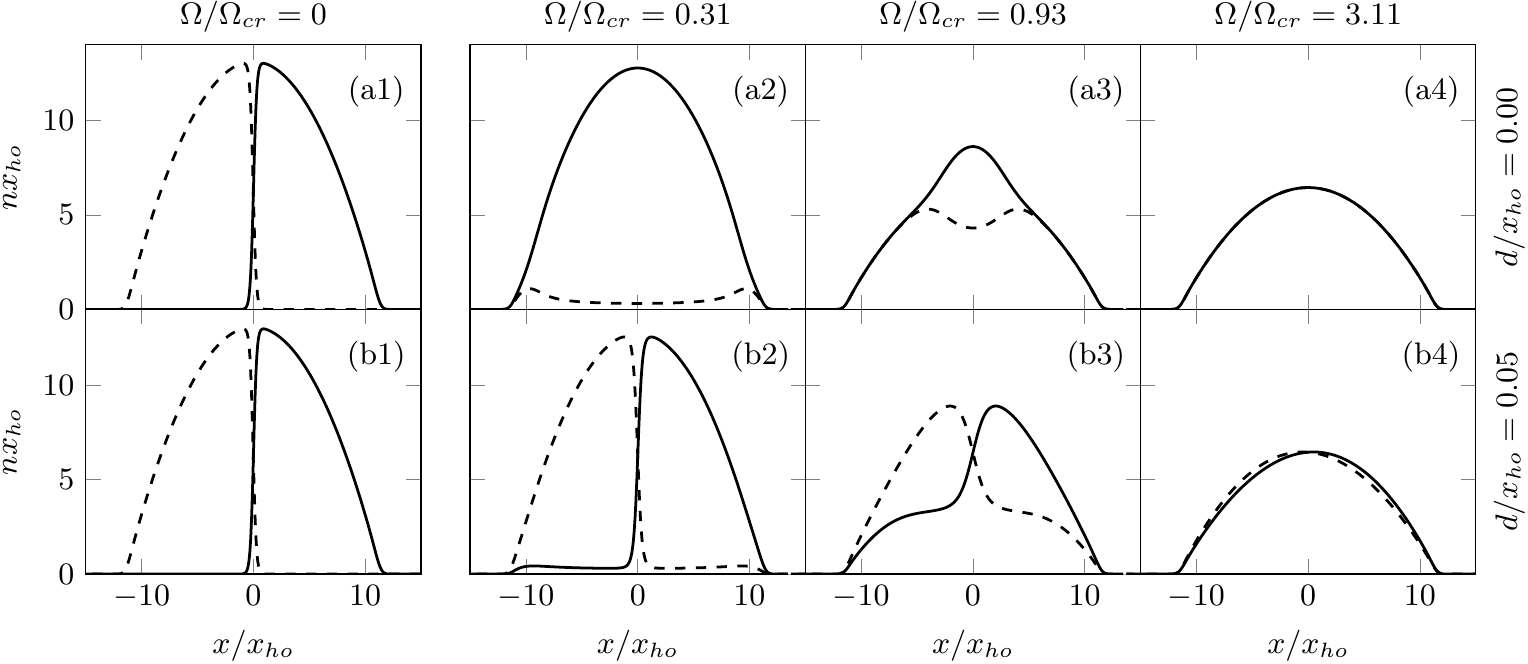}
\caption{Ground states of a trapped coherently-coupled Bose gas for different values of potential displacement $d$ and Rabi coupling $\Omega$. We use coupling constant strength $g_{ab}/g=1.1$ and $g/(\hbar\omega_{ho}x_{ho}) = 5$. Plots (a1)-(a4) correspond to $d/x_{ho}=0$, while plots (b1)-(b4) to $d/x_{ho}=0.05$. For the ground state in panels 2, 3, 4 we use the values $\Omega/\Omega_{cr}=0.31,\ 0.93,\ 3.11$, respectively. For comparison we report also the ground state for a Bose-Bose mixture, i.e., $\Omega=0$. In the latter case, being the number of particle in each component fixed, no global polarisation appears, and the ground state (a1) and (b1) are essentially equal. The effect of quantum pressure can be clearly  noticed in the plots (a2) and (a3) (analogues to Fig.~\ref{fig:phase}).  The  bifurcation points are not sharp as instead predicted by the Thomas-Fermi approximation.}\label{fig:dipgs}
\end{figure*}


The GP ground state solution for the spinor gas in the displaced potentials is reported in Fig.~\ref{fig:dipgs} (see also \cite{Blakie1999}), where for the sake of concreteness we assume $g_{ab}>g$ to show the difference between a mixture and a coherently driven spinor gas. For $d=0$ (row a)  we see the features of the $\Omega$-induced phase transition: below the critical value the linear coupling prevents the phase separation by creating a global polarisation in the system (Fig.~\ref{fig:dipgs} plots (a2) and (a3)), while a mixture without any Rabi coupling $\Omega=0$ would be in a phase separated state (Fig.~\ref{fig:dipgs} plots (a1)). Above the critical value the gas is unpolarised (Fig.~\ref{fig:dipgs} plot (a4)). Applying a potential displacement (row b) makes the local polarisation different from zero as shown in Eq. (\ref{eq:gs1}). In this case even a small potential difference makes the ferromagnetic part of the gas strongly polarised and as a result a magnetic domain wall is created at the center of the trap.

In order to calculate the spin-dipole susceptibility we first determine the spin-dipole moment $D$, defined as
\be
D = \frac{1}{N}\int x\left[n_a(x)-n_b(x)\right]\mathrm{d}x.\label{eq:D}
\ee
The spin-dipole susceptibility is then defined by the limit
\be
\chi_{sd}=\lim_{d\rightarrow 0} D/\lambda
\label{chilim}
\ee
where $\lambda = dm\omega^2_{ho}$ is the perturbation associated with the spin-dependent component of the potential (\ref{eq:diptrap}).

In the {\sl global paramagnetic phase} ($\Omega> \Omega_{cr}$) it is easy to obtain an analytical expression for $\chi_{sd}$ within LDA. In this case one can employ the energy functional
\begin{equation}
E= \int \left[ \chi_s^{-1} (n_a-n_b)^2 -\lambda x(n_a-n_b)\right]\;dx
\label{ELDA}
\end{equation}
relative to the spin degrees of freedom of the problem, where
\be
\chi_s = \frac{2}{g-g_{ab}+\Omega/n_0}
\label{eq:chi}
\ee
is the spin (magnetic) susceptibility for an homogeneous system of density $2n_0$ (see, e.g., \cite{marta2013}). Variation with respect to the spin density $(n_a-n_b)$
yields the result
\be
n_a(x)-n_b(x)=x\lambda\chi_s(n_0(x)),
\ee  
and the spin dipole polarizability finally reads 
\be
\chi_{sd}=\frac{1}{N}\!\int\! x^2\chi_s(n_0(x)).
\label{chisd}
\ee
After integratiion we obtain the result
\be
\frac{D}{d}=\frac{g+g_{ab}}{g-g_{ab}}\left[\! 1\!+f\left(\frac{\Omega}{(g-g_{ab})n_0}\right)\right],
\label{eq:sdmom}
\ee
for the dimensionless ratio $D/d=m\omega_{ho}^2\chi_{sd}$ where we have  introduced the dimensional function 
$f(\alpha)=  3\alpha\left(1 -\sqrt{1+\alpha}\;\mathrm{arccoth}(\sqrt{1+\alpha})\right)$ \footnote{Notice that the domain of the function $f(\alpha)$ (to be real),  i.e., $1/\alpha\ge -1$ is precisely where the system is fully paramagnetic}  and used the notation $n_0=n_0(0)$. 

\begin{figure}[t]
\centering
\includegraphics{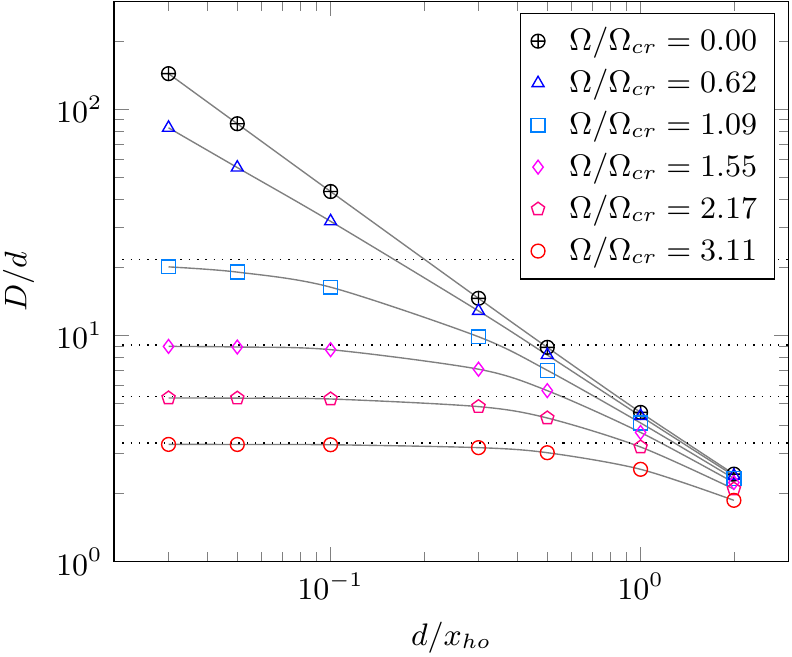}
\caption{Dipole $D$ as a function of traps displacement $d$ for different values of $\Omega/\Omega_{cr}$ and for $g_{ab}/g=1.1$ as in Fig. \ref{fig:dipgs}. Dotted lines are analytical results from eq. (\ref{eq:sdmom}) for the four bigger values of $\Omega$, points are numerical data and grey full lines are only a guide for the eyes.}\label{fig:Dd}
\end{figure}

A couple of comments are due here.
First of all let us consider the case of a Bose-Bose mixture, i.e., $\Omega \rightarrow 0$. In this case $f(\alpha\rightarrow 0)\rightarrow 0$ and the spin-dipole susceptibility is simply proportional to the magnetic susceptibility Eq. (\ref{eq:chi}). Therefore also $\chi_{sd}$ diverge at the (miscible-immiscible) transition point $g_{ab}\rightarrow g^-$ \footnote{At the same point, but for finite $\Omega$, one has $f(\alpha\rightarrow\infty)\rightarrow -1+2/(5\alpha)$ and therefore $D/d=gn_0/(5\Omega)$.}. 
Physically this is due to the fact the two gases become globally immiscible at the transition point since the latter condition is density independent.
As we will see in the next Section the divergence of $\chi_{sd}$ leads to a zero frequency (soft) spin-dipole mode. 

On the  other hand for finite $\Omega$ the paramagnetic-ferromagnetic transition point, namely $g_{ab}=g+\Omega/n_0$, depends on the density. 
Therefore the spinor gas starts becoming ferromagnetic at the center of the trap only. The quantity $\chi_{sd}$, being density integrated, remains finite at the transition point  (indeed $f(-1)=-3$) leading (see next Section) to a finite frequency for the spin-dipole mode. This behaviour is very general and it has already been pointed out for the Stoner (or itinerant ferromagnetic) instability in the context of cold gases by two of us \cite{recati2011}.

Above the critical point the response of the system is no longer linear. The system is {\sl partially ferromagnetic} and has the tendency to form a magnetic domain wall at the centre of the trap (see Appendix). 

A detailed analysis of the behaviour of $D/d$ is shown in Fig.~\ref{fig:Dd} where we calculate numerically the spin-dipole of the gas as a function of the trap separation $d$ with the choice $g_{ab}/g=1.1$. Above the critical Rabi frequency we see that indeed linear response applies and the analytical expression Eq. (\ref{eq:sdmom}) works very well. Notice that the spin-dipole moment allows for a clear  identification of  the phase transition point, above which the induced dipole moment $D$ changes its behavior  as a function of $d$.

\section{Spin dipole dynamics}

In this section we study the dynamics of the system. 
In particular we prepare the system initially in the ground state of very slightly displaced external potential and then suddenly set the displacement to zero.
As one can expect the physics is completely different depending on whether the system is paramagnetic or ferromagnetic. 
In the earlier case the system shows a well defined out-of-phase oscillation, the spin-dipole mode. The previously calculated spin-dipole polarizability plays an important role in characterising the behavior of the spin-dipole  frequency \cite{recati2011,SandroSO-collective}. Notice in particular that for two independent condensates ($\Omega=g_{ab}=0$) the spin-dipole frequency would simply coincides with the trap frequency $\omega_{ho}$.
In the ferromagnetic case the system evolves according to a highly non-linear dynamics and it shows ground state selection. 
We analyse the two cases separately in the next two sections. Some details on the numerical solution of the GP equations can be found in Appendix B and reference therein. For the interested reader we include in the Supplementary Material the real time evolution of the system in different regimes.   

\subsection{Paramagnetic phase: sum rule approach}\label{sec:para} 

In the paramagnetic phase, as shown in Fig. (\ref{fig:dipgs}) (a4)-(b4), a small trap displacement corresponds to a small deviation with respect to the ground state at zero displacement and therefore linear response theory can be applied.  The dynamics we consider in this case coincides with the dynamical response of the spinor gas to the spin-dipole operator $\hat{S}_d=\sum_{i=1}^N x_i\hat{\sigma}_{z,i}$. A very powerful tool to estimate the frequency of collective modes is the so-called sum rule approach  \cite{Bohigas,SandroLippa}. This approach has been very successfully employed for the dynamics of both cold gases and nuclei. 
We remind here simply that sum rules are defined for an operator $\hat{F}$ as
\be
m_k(F) = \sum_n |\langle0|\hat{F}|n\rangle|^2(E_n-E_0)^k,
\ee
and they represents the moments of the strength distribution function relative to $\hat{F}$.
The sum rule approach has the merit of providing a direct way to obtain an upper bound of collective mode frequency through the ratio of different sum rules, and therefore giving an understanding of the collective mode frequency in terms of static macroscopic quantities  \cite{SandroLippa}. 

In our case the operator of interest is $\hat{S}_d$ and we we use the energy weighted and inverse energy weighted sum rule, i.e.,  
\be
\hbar^2{\omega_{\mathrm{SD}}}^2 \leq \frac{m_1(S_d)}{m_{-1}(S_d)}.
\label{sumrule}
\ee

They are particularly suited in our case. The energy weighted one ($m_1$) is easily rewritten in terms of a double commutator as $m_1=(1/2)\langle0|[S_d,[H,S_d]]|0\rangle$. The only terms in $H$ that do not commute with $S_d$ are the kinetic energy and the  Rabi coupling $H_R=-\Omega \sum_i\hat{\sigma}_{x,i}$. The former gives the usual $N\hbar^2/(2m)$ contribution, while the latter is straightforwardly evaluated as $-4\Omega x^2\hat{\sigma}_x$. Averaging on the ground state, we obtain
the result
\be
m_1 = N\frac{\hbar^2}{2m} +8\Omega\int_0^{\rtf} x^2 n_0(x) \dd x.
\label{eq:frule}
\ee
The inverse energy weighted sum rule ($m_{-1}$) is directly related to the susceptibility of the ground state through the relation 
\be
m_{-1} = \frac{N}{2}\chi_{sd},
\ee
and using the definition Eq. (\ref{chilim}) together with the result Eq. (\ref{chisd}) we obtain the following upper bound to the spin-dipole frequency 
\be
{\omega_{\mathrm{SD}}}^2 = \ohosq \left(\frac{g-g_{ab}}{g+g_{ab}}\right) \left[\frac{1+8\Omega n_0(g+g_{ab})/(5\hbar\omega_0^2)}{1+f(\Omega/((g-g_{ab})n_0))}\right].\label{eq:osd}
\ee

Notice that the equality in Eq. (\ref{sumrule}) is attained when the whole strength is exhausted by a single state.

\begin{figure}
\centering
\includegraphics{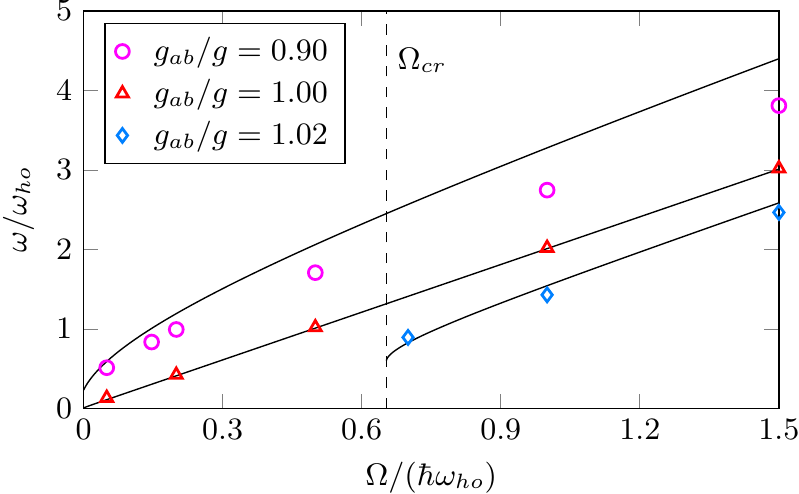}
\caption{Spin-dipole frequency as a function of $\Omega$ for different values of interactions. Lines are analytical results from Eq.~(\ref{eq:osd}) and points are numerical data. In order to have a fully paramagnetic phase for $g_{ab} > g$ one needs $\Omega \ge \Omega_{cr}$ (see text and Eq. (\ref{omegacr})), the value of $\Omega_{cr}$ for $g_{ab}/g=1.02$ (blue diamonds) is shown by the dashed line. In the Supplementary Material, two videos show the oscillations of the clouds for $g_{ab}/g=0.9$ and $g_{ab}/g=1$ (both at $\Omega=0.5\hbar\omega_{ho}$): $1.0$ sec of the video corresponds to $1.0\omega_{ho}t$.}\label{fig:osd}

\end{figure}

In Fig.  \ref{fig:osd} the sum-rule result is compared with the predictions of the solutions of  a time dependent  Gross-Pitaevskii calculation. As already mentioned from the numerical or experimental point of view, the excitation of  the spin-dipole mode is achieved starting with  an equilibrium  configuration in the presence of slightly displaced trapping potentials, as described by Eq.~(\ref{eq:diptrap}), and  suddenly setting $d=0$. 

\begin{figure}
\centering
\includegraphics{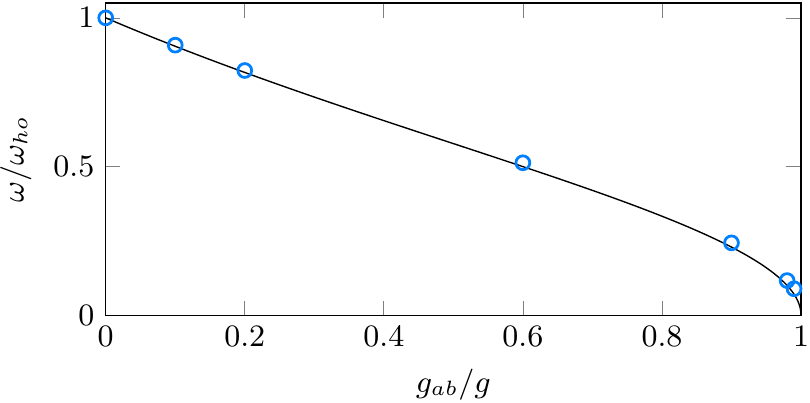}
\caption{Spin-dipole frequency for a Bose-Bose mixture, i.e. $\Omega=0$, as a function of the ratio $g_{ab}/g$. Line is the analytical result and points are numerical data. In this case the spin-dipole frequency goes to zero at the phase separation transition point.}\label{fig:osdmix}
\end{figure}

Notice that at the transition point the frequency does not go to zero, since for the reasons explained in the previous Section $\chi_{sd}$ (or $m_{-1}$) does not diverge at that point. This has to be compared with the mixture case, which is recovered sending $\Omega\to0$. In this case the spin-dipole frequency vanishes close to the critical point following the law 
\be
\omega_{SD}(\Omega=0)= \omega_{ho}\sqrt{\frac{g-g_{ab}}{g+g_{ab}}} \; ,
\label{omegaSDOmega=0}
\ee
and the sum-rule approach give the exact result as shown in Fig. \ref{fig:osdmix}.

Sum-rules give the exact result also for the intrinsic $SU(2)$ symmetric point $g_{ab}=g$ (and $\Omega\neq 0$ in general)and $g_{ab}=g$ as it can be seen in Fig. \ref{fig:osd}) (red triangles). 
The magnetic energy of the spinor gas in this regime depends on the relative density only through the Rabi coupling which breaks the $SU(2)$ symmetry of the system. The spin-dipole  frequency behaves in this case  as 
\be
\omega_{\mathrm{SD}} (g_{ab}=g) = 2\Omega\sqrt{1+\frac{5}{16}\frac{\hbar\oho^2}{gn_0 \Omega}}
\ee
which is essentially twice the Rabi frequency and therefore almost independent of the tapping frequency.
The latter unusual result for a trapped gas is due to the correlation between the internal and external degrees of freedom that in particular lead to the modification of the $f$-sum rule, see Eq. (\ref{eq:frule}).

In the more general case, when both $\Omega$ and $(g_{ab}- g)$ are different from zero, the frequency is given by the full eq. (\ref{eq:osd}) in which both the coherent and the interspecies $s$-wave couplings play a role. In this more general case one observes that that the sum rule approach provides only an upper bound to the numerical solution, due to the appearance of more frequencies in the numerical signal resulting in beating effects. 

In the Supplementary Material we include two videos showing the oscillations of the clouds in the paramagnetic case for $g_{ab}/g=0.9$ and $g_{ab}/g=1$ and $\Omega=0.5\hbar\omega_{ho}$. The real time evolution shows clearly the presence of only one frequency in the intrinsic $SU(2)$ symmetric case and the appearance of more frequencies when the $g_{ab}\neq g$.

\begin{figure*}[!t]
\centering
\includegraphics{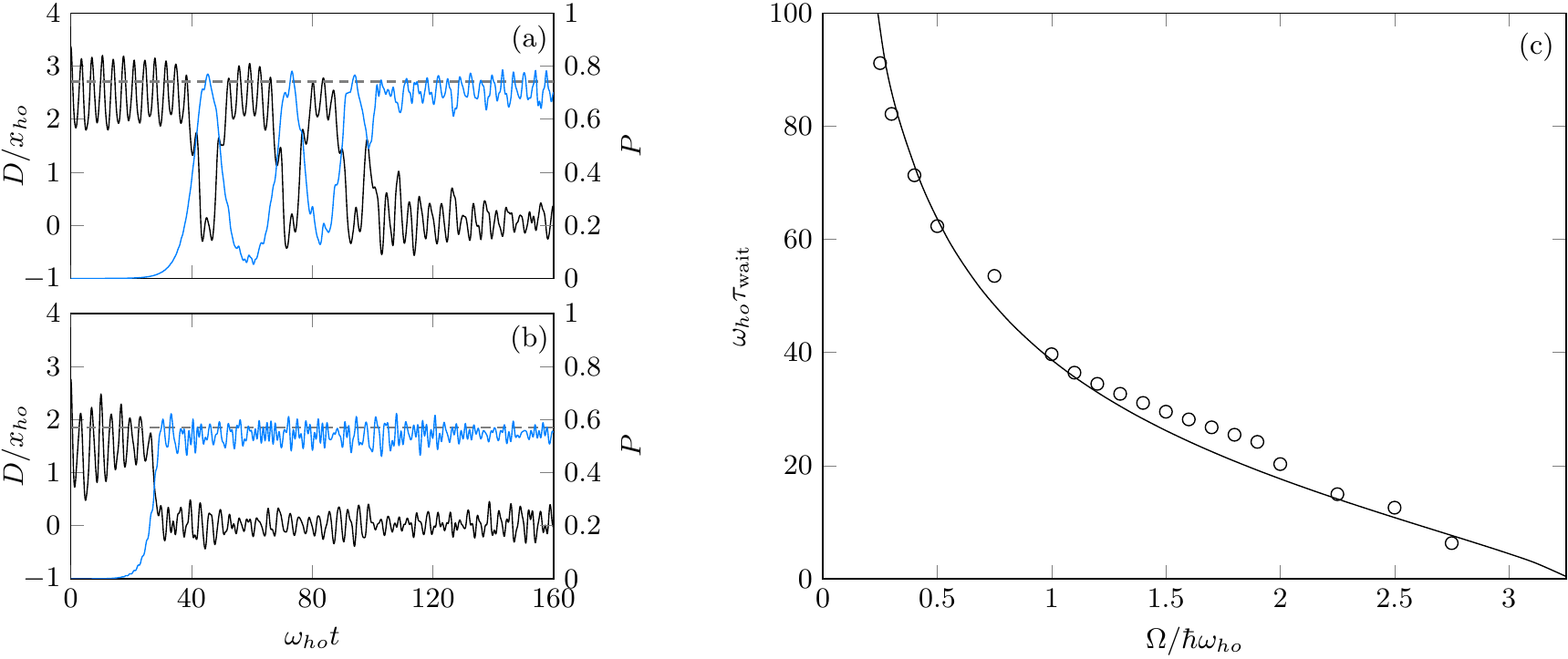}
\caption{(a) and (b): real time spin-dipole oscillation (black solid lines), polarization of the system (blue solid lines) and ground state polarization values (dashed lines); parameters are $g_{ab}=1.1g$, $g/(\hbar\omega_{ho}x_{ho}) = 5$ and $\Omega/\hbar\omega_{ho}=1.5$ (a), $\Omega/\hbar\omega_{ho}=2$ (b). In the Supplementary Material a video shows the oscillation of the clouds of panel (a), $1.0$ sec of the video corresponds to $1.0\omega_{ho}t$.  (c): waiting time as a function of $\Omega$, points are numerical data and line is a fit of data with function $A\sqrt{\sigma}$ where $\sigma$ is the surface tension of Eq.~(\ref{tens}). } \label{fig:dipdin}
\end{figure*} 

\subsection{Ferromagnetic phase: ground state relaxation}\label{sec:ferro}

In the previous section we have  studied the dynamics for a completely paramagnetic gas, i.e., $\Omega>\Omega_{cr}$. The behaviour is very different when the system presents a ferromagnetic behavior. In this case the ground state of the system with equal trapping potentials is polarized as shown in Fig.~\ref{fig:dipgs} plots (a2) and (a3). When the traps are shifted, the ground state is instead  globally unpolarized   ($N_a = N_b$) but with a large spin-dipole moment (depending on the values of $\Omega$ and $d$) as one can see in Fig.~\ref{fig:dipgs} plots (b2) and (b3). Therefore the initial state and the ground state are very  far from each other. This circumstance results in a non-trivial non-linear dynamics as shown by the dynamics of the spin-dipole and of the polarisation reported in Fig. \ref{fig:dipdin}.
At the beginning, the spinor gas oscillates around the initial configuration, trapped in the unpolarized state. After a certain time, $\tau_\mathrm{wait}$, the domain wall starts moving and a finite polarisation appears. The system then bounces back and forth between the initial magnetic state and its magnetic ground state  to eventually relax to the latter one. \footnote{The system is closed and energy conserving and still able, in the long time limit, to approach and select one of the two possible ground states. The final state obviously presents still (small) oscillations around its ground state.} An example of such dynamics can be viewed in the video in the Supplementary Material. If the global polarisation of the ground state is large, the effects of non linearity and the number of  bounces are large. When the system is slightly in the ferromagnetic regime no bounces are observed and the system after $\tau_\mathrm{wait}$ soon reaches its ground state (see right lower panel in Fig. \ref{fig:dipdin}). 
%
Notice that even if the system is isolated, it can approach in the long time limit an asymptotic steady state, as a result of destructive interference of several time oscillating  factors, present in the evolution of expectation values of observables. Specifically, in the case of a large and dense collection of frequencies, the interference phenomenon results in a dephasing mechanism similar to inhomogeneous dephasing. 

The total energy of the system is still conserved, the algorithm used (see Appendix \ref{numerics}) does not contains any dissipative mechanism and we explicitly check that the total energy does not change during the evolution. At the end of the real time evolution we get the ground state profile with superposed some high frequency perturbations carrying the extra energy.

As we already mentioned in Sec. II the initial configuration in the ferromagnetic case contains a domain wall at the centre of the trap. We have identified a close relation between the observed waiting time and the square root of the domain wall energy (see Appendix \ref{appendix})
\be\sigma\propto{\frac{|(g-g_{ab}) n +2\Omega|^{3/2}}{\Omega}}\label{tens}.\ee 
From an intuitive point of view, the higher is the energy of the domain wall, $\sigma$, the more time is required to the system to relax from the kink into one of the ground states of the system; accordingly, it is expected a relation of proportionality between the waiting time and $\sigma$. A standard field theoretical estimate of the average tunnelling time cannot be straightforwardly performed since only close the transition our field theory resembles an ordinary $\phi^4$ theory (see Appendix A for details); for this reason, we benefited of a numerical fit to extract with surprising accuracy the relation, $\tau_{wait}\propto\sqrt{\sigma}$, as shown in Fig. \ref{fig:dipdin}.

The fact that for $\Omega\rightarrow 0$ the waiting time diverges can be easily understood noticing that the initial state and the ground state are very far from each other (see, e.g., panels (a2) and (b2) of Fig.~\ref{fig:dipgs}). Eventually, in the strict $\Omega=0$ case, the system cannot reach the totally polarised ground state and it remains in the phase separated state (see panels (a1) and (b1) of Fig.~\ref{fig:dipgs}).

\section{Conclusions}

In the present work we analyse in details the static and dynamic response of a trapped coherently driven 2-component condensate to spin-dipole probe. We show that the spin-dipole susceptibility is a good quantity able to identify the appearance of a ferromagnetic-like region in the cloud. 

For the dynamics we study the spin-dipole mode frequency by starting in a configuration with displaced harmonic potentials, which are suddenly brought to the same value. When the system is paramagnetic such a frequency is well reproduced by a sum-rule approach. In particular the $f$-sum rule is strongly modified by the Rabi coupling in the symmetric interaction case ($g=g_{ab}$) and the inverse energy sum rule is proportional to the spin-dipole susceptibility and coincides with the second spatial momentum of the local magnetic susceptibility (see Eq.(\ref{chisd})).

When the system has a ferromagnetic domain a linear response cannot be applied anymore and the dynamics is highly non-linear. The initial configuration within displayed potentials is unpolarised and contains a magnetic kink centred at the origin.  The dynamics is trapped for a time, $\tau_\mathrm{wait}$, in the initial configuration, after which the system is able to relax to its polarised ground state. We find that $\tau_\mathrm{wait}$ is proportional to the square root of the kink surface tension.

Our study improves the characterisation of coherently driven BECs, enlightening their differences with respect to Bose-Bose mixtures. Moreover measuring the spin-dipole dynamics  opens new perspective to  experimentally access important magnetic properties of the system, as, e.g., its susceptibility or the domain wall surface tension.  

\appendix
\section{Magnetic domain wall surface tension}\label{appendix}


In this Appendix we briefly show how to approximate the energy functional for the magnetisation for a spinor condensate with a classical one dimensional $\phi^4$ (Ginzburg-Landau for the phase transition) theory  \cite{LL5}. From the latter it is easier to show the existence of a kink or domain wall in the magnetization, and we compute in this regime its surface tension. 
In the symmetric case $g_a=g_b=g$ and considering an uniform total density  $n=n_a+n_b$, the relative density or magnetisation $M=(n_a-n_b)/n$, enters 
in the energy density 
\be
E(M)=\int dx \left[\frac{\hbar^2 n(\nabla{M})^2}{8m\left(1-M^2\right)}+W(M)\right],
\label{em}
\ee
where the first term arises from the kinetic energy and the term 
\be
W(M)=\frac{n^2}{4}(g-g_{ab})M^2-\Omega n \sqrt{1-M^2},
\label{WM}
\ee
accounts for the density-density interaction and the Rabi terms. 
For a homogeneous magnetization minimisation $\delta E/\delta M=0$ leads to the usual equation for the para- and ferro-magnetic like states.
From Eq. (\ref{em}) one sees that close to the phase transition, i.e., $M\ll 1$ a standard Ginzburg-Landau theory for the order parameter $M$, is valid, where the kinetic energy is just the the square of the gradient of $M$ and the effective potential  takes the usual quadratic plus quartic form
\begin{align}
\nonumber W(M) &=\frac{n^2}{4}\left(g-g_{ab} +\frac{2\Omega}{n}\right)M^2+\frac{n\Omega}{8} M^4\\
&\equiv\frac{r}{2} M^2+ \frac{u}{4} M^4.
\end{align} 
As usual the $\mathbb Z_2$ symmetry broken ground state is obtained for $r<0$. 
A kink in $M$ is the field solution interpolating between the two degenerate minima. Its surface tension, $\sigma$, which coincides with its energy  in a one-dimensional situation, can be easily computed \cite{WeimbergClassical} yielding the result 
\be
\sigma\propto \sqrt{\frac{\hbar^2 n^2}{m}} {\frac{|r|^{3/2}}{u}}\propto  \sqrt{\frac{\hbar^2 n^2}{m}}{\frac{|\delta g n +2\Omega|^{3/2}}{\Omega}}
\label{sigmaru}
\ee
 

\section{Numerical method}\label{numerics}

All numerical data presented in this paper have been obtained solving the GP coupled equations by means of the split-operator method and by treating the kinetic term in Fourier space \cite{numBao,numJaksch}. The initial wave functions $\psi_a(x,t)$  and $\psi_b(x,t)$ are evolved for a time step $\Delta t$ alternately by the kinetic, potential and Rabi terms of Hamiltonians in Eq. (\ref{eq:tGP}):

\begin{align}\label{eq:evolution}
\tilde{\psi}_i(k,t)\mapsto& \,\mathrm{e}^{-ik^2\Delta t/2}\tilde{\psi}_i(k,t)\nonumber\\
\psi_i(x,t) \mapsto& \,\mathrm{e}^{-i(V_i + g|\psi_i|^2 +g_{ab}|\psi_j|^2)\Delta t}\psi_i(x,t)\\
\psi_i(x,t) \mapsto& \,\mathrm{cosh}(-\Omega \Delta t)\psi_i(x,t) + \mathrm{sinh}(-\Omega \Delta t)\psi_j(x,t)\nonumber
\end{align}
where $i=a,b\neq j$ and Eq. (\ref{eq:evolution}) is for the imaginary time evolution. One can simply obtain the same set of equations for real time evolution by changing $\Delta t$ in $\ii\Delta t$.

An algorithm of this type is symplectic, this means that the method exactly simulates a Hamiltonian $H_{\Delta t}$ with $H_{\Delta t} - H$ a power series in $\Delta t$. Advantages of using symplectic integrators are that there is no drift in energy due to the exact conservation of $H_{\Delta t}$ and that the phase-space volume is exactly conserved.

In order to obtain the ground states we ran the imaginary time evolution starting from an initial trial wave function built from both random density and phase distributions, in order to prevent the algorithm from reaching false metastable states. For the dynamics we loaded the ground states obtained with displaced traps and let them evolve using the same algorithm but with real time and with the equal trapping potential Hamiltonian. The values of polarization $P$ and of spin-dipole moment $D$ are calculated and saved at each time step and then analysed to obtain the frequencies. This last step is not always so straightforward, sometimes the signal contains more than one frequency and damping occurs. In this cases we perform a Fourier analysis of the data and we keep the maximum-amplitude frequency.

\acknowledgments

A.R. acknowledges useful discussions with Marta Abad, Markus Heyl, Nicolas Pavloff, and Wilhelm Zwerger and support from the Alexander von Humboldt foundation. 
A.S. acknowledges Komet 337 group of the Johannes Gutenberg Universit\"at Mainz for the kind hospitality. This work has been supported by ERC through the QGBE grant and by Provincia Autonoma di Trento. 

\bibliography{NJP-sub2_02072015}{}

\end{document}